\documentclass[pdflatex,sn-mathphys-num,iicol]{sn-jnl}

\usepackage{graphicx}%
\usepackage{multirow}%
\usepackage{amsmath,amssymb,amsfonts}%
\usepackage{amsthm}%
\usepackage{mathrsfs}%
\usepackage[title]{appendix}%
\usepackage{xcolor}%
\usepackage{textcomp}%
\usepackage{manyfoot}%
\usepackage{booktabs}%
\usepackage{algorithm}%
\usepackage{algorithmicx}%
\usepackage{algpseudocode}%
\usepackage{listings}%

\begin{document}

\title[AI Grading Assistance of a Calculus Exam]{Artificial-Intelligence Grading Assistance for Handwritten Components of a Calculus Exam}


\author*[1,2]{\fnm{Gerd} \sur{Kortemeyer}}\email{kgerd@ethz.ch}

\author[3]{\fnm{Alexander} \sur{Caspar}}\email{caspara@ethz.ch}

\author[1]{\fnm{Daria} \sur{Horica}}\email{donishchuk@ethz.ch}

\affil*[1]{\orgdiv{Rectorate and ETH AI Center}, \orgname{ETH Z{\"u}rich}, \orgaddress{\street{R{\"a}mistrasse 101}, \city{Z{\"u}rich}, \postcode{8092}, \country{Switzerland}}}

\affil[2]{\orgname{Michigan State University},  \orgaddress{\city{East Lansing}, \postcode{48824}, \state{Michigan}, \country{USA}}}

\affil[3]{\orgdiv{Department of Mathematics}, \orgname{ETH Z{\"u}rich}, \orgaddress{\street{R\"amistrasse 101}, \city{Z{\"u}rich}, \postcode{8092}, \country{Switzerland}}}


\abstract{We investigate whether contemporary multimodal LLMs can assist with grading open-ended calculus at scale without eroding validity. In a large first-year exam, students' handwritten work was graded by GPT-5 against the same rubric used by teaching assistants (TAs), with fractional credit permitted; TA rubric decisions served as ground truth. We calibrated a human-in-the-loop filter that combines a partial-credit threshold with an Item Response Theory (2PL) risk measure based on the deviation between the AI score and the model-expected score for each student-item. Unfiltered AI-TA agreement was moderate, adequate for low-stakes feedback but not for high-stakes use. Confidence filtering made the workload-quality trade-off explicit: under stricter settings, AI delivered human-level accuracy, but also left roughly 70\% of the  items to be graded by humans. Psychometric patterns were constrained by low stakes on the open-ended portion, a small set of rubric checkpoints, and occasional misalignment between designated answer regions and where work appeared. Practical adjustments such as slightly higher weight and protected time, a few rubric-visible substeps, stronger spatial anchoring should raise ceiling performance. Overall, calibrated confidence and conservative routing enable AI to reliably handle a sizable subset of routine cases while reserving expert judgment for ambiguous or pedagogically rich responses.}

\keywords{Artificial Intelligence, Assessment, Mathematics Education}

\maketitle

\section{Introduction}\label{sec:intro}

Calculus courses serve as gateways to advanced study in virtually all STEM disciplines. In these courses, students are expected not only to obtain correct numerical or algebraic results but also to formulate models, select appropriate techniques, justify steps, and communicate reasoning with symbols, diagrams, and words. Research on learning and assessment consistently shows that such open-ended work --- showing the path, not just the endpoint --- better reflects the knowledge and practices we intend to cultivate, supports transfer, and makes students' conceptions visible for feedback and instruction~\cite{bloom56,laverty12b,offerdahl2019formative}. Yet in large-enrollment settings, logistical pressure often pushes assessment toward closed-answer formats that can be machine-graded at scale, narrowing what is measured and learned.

Closed-answer technologies have a long history, from mechanical multiple-choice devices to modern web-based systems that grade numeric responses with tolerances, check algebraic equivalence, and match short strings~\cite{petrina2004sidney,suppes1966arithmetic,sangwin2007assessing,kortemeyer08}. These approaches can deliver high scoring consistency when the space of correct answers is well defined. However, they struggle to capture the multi-step reasoning, representation shifts, and communication quality that matter in mathematics learning, especially in calculus where procedures (e.g., differentiation techniques) intertwine with concepts (e.g., limits, continuity, and the meaning of the derivative) and with modeling and interpretation~\cite{teodorescu2013new,meltzer2005relation}. Automatic Short Answer Grading (ASAG) occupies a middle ground by recognizing paraphrase and semantic equivalence in brief textual responses, typically with machine-learning models~\cite{leacock2003c,zhang2022automatic,ahmed2022deep}. Yet ASAG generally does not address the mixed-media nature of authentic calculus work: symbolic derivations, structured chains of reasoning, and sketches of graphs or geometric configurations.

Recent advances in large language models (LLMs) and multimodal systems have re-opened the possibility of grading assistance for open-ended work at scale~\cite{chatgpt,gpt4,gpt4o}. Across various fields of education, LLMs can evaluate and generate academic work and support research workflows~\cite{meyer2023chatgpt,kasneci2023chatgpt,tschisgale2023integrating,kieser2023educational,tolstykh2024beyond,campbell2025applying}; in STEM education, researchers have begun to explore their use for solving, generating, and assessing problems~\cite{yeadon2024impact,sperling2024artificial,polverini2024understanding,kortemeyer23ai}. Our own prior studies suggest both promise and limits. First, benchmarking GPT-4 on ASAG showed performance comparable to earlier hand-engineered systems, with the practical advantage of no task-specific training and, in some cases, grading without reference solutions~\cite{kortemeyer2024performance}. Second, in handwritten undergraduate mathematics, we proposed a framework to estimate the reliability of short-answer grades and found that recognition of mathematical notation is a key bottleneck; scanning and transcribing whole pages before extracting answers can mitigate errors~\cite{liu2024ai}. Third, in open-ended physics and chemistry exams, end-to-end pipelines revealed that layout and recognition quality strongly affect downstream grading, and that human-in-the-loop routing remains essential due to occasional high-confidence errors~\cite{kortemeyer24aigrading,kortemeyer2024grading,cvengros2025assisting}. These studies also demonstrated the value of psychometric instrumentation, using Item Response Theory (IRT) and related tools to quantify when and how AI grades can be trusted~\cite{lord2008statistical,pawl2013,scott2006,kortemeyer2016psy,kortemeyer2025assessing}.

A central technical and methodological challenge for mathematics education is that student work is genuinely mixed-media. Mathematical handwriting must be transcribed with structure preserved; diagrams and graphs need faithful descriptions; and printed headers and rubric anchors must remain aligned through scanning and registration. Classical OCR is strained by mathematics and layout variability~\cite{mori1992historical,okamura1999handwriting,wang2021}; while specialized tools can help for formulas~\cite{mathpix}, multimodal LLMs promise more integrated ``see-and-grade'' pathways~\cite{gpt4o}. At the same time, educational use requires calibrated uncertainty: instructors need to know when automation is reliable without first grading everything by hand. Human-in-the-loop designs, i.e., routing clear, routine cases to automation and flagging uncertain or pedagogically rich cases for expert review, offer a pragmatic compromise~\cite{de2020case,alsalmani23a}. To be trustworthy, such systems must report well-calibrated confidence, meet explicit quality targets, and support appeals and transparency obligations that many jurisdictions now expect for high-stakes educational AI~\cite{euactannex,euactobl}.

We are studying AI assistance for grading the open-ended portions in a large-enrollment, first-year university calculus exam. Due to the above-mentioned constraints, 2/3rds of this exam is closed-answer, but four problems were left open-ended. We focus on three intertwined questions. First, to what extent can contemporary AI systems, used within a human-in-the-loop workflow, support reliable evaluation of open-ended calculus responses at scale, including symbolic derivations and brief written justifications? Second, how should confidence be quantified and calibrated so that auto-accepted decisions meet conservative precision targets while ambiguous or novel cases are efficiently routed to human graders? Third, what aspects of exam layout and recognition (e.g., page anchors, boxed answer regions, transcript quality) materially affect agreement with expert grades?

Our design draws on prior evidence and seeks mathematics-specific insight. Building on our ASAG benchmark~\cite{kortemeyer2024performance} and reliability framework for handwritten mathematics~\cite{liu2024ai}, we treat confidence calibration and routing as first-class objects: we integrate model-based signals (e.g., log probabilities, self-consistency), rubric alignment, and transcript/recognition quality into a calibrated probability of correctness, then set operating points that respect course and departmental constraints~\cite{kortemeyer2025assessing}. In parallel, we consider practical layout affordances linked to recognition and rubric alignment, echoing findings from physics and chemistry that small design choices can have outsized downstream effects~\cite{kortemeyer2024grading,cvengros2025assisting}. Throughout, we frame validity in terms of what matters for calculus learning: visibility into reasoning, representation changes (algebraic steps, graphs), and communication quality, not only terminal answers~\cite{meltzer2005relation,teodorescu2013new}.

Taken together, the study advances a use-inspired, evidence-grounded approach to AI-assisted grading in mathematics education. Rather than replacing expert judgment, we aim to reserve it for the cases where it is most needed by combining (i) authentic student work on paper, (ii) recognition and multimodal analysis tuned to mathematical structure, and (iii) psychometric calibration of uncertainty and decision thresholds. In doing so, we address a practical constraint in large-enrollment calculus while preserving the educational value of assessing how students think, justify, and communicate in mathematics.

\section{Methods}\label{sec:methods}
\subsection{Exam}
The study was conducted in a year-long mathematics course for biology, chemistry, and health science students taught by one of the authors (A.~C.). The exam covered both semesters. Because of limited grading staff, most items were closed-answer: a multipart, multiple-choice problem~1 accounted for most points. Four problems (2--5) were open-ended; problem 5 had two parts. In particular, these open-ended problems are:

\begin{itemize}

\item {\bf 2.A1:}
The problem deals with a parameterized $3\times3$ matrix and the interplay between determinant, rank, and solvability of a homogeneous linear system. A challenge is to express $\det(D_b)$ as a function of the parameter and to translate $\det(D_b)\neq0$ into an invertibility statement, while recognizing the exceptional value that yields a nontrivial kernel. The student needs to master basic determinant techniques (e.g., Sarrus/Laplace), recognize that invertibility hinges on the determinant, and be able to characterize the nullspace (rank-nullity, dimension of the solution set) in the singular case.

\item {\bf 3.A1:}
The problem deals with a first-order separable differential equation and the global domain of its solutions. A challenge is to separate variables cleanly, integrate, and handle the constant of integration so that the resulting family $y(x) = -1/(x^2+C)$ is interpreted correctly with respect to the initial value and possible singularities. The student needs to master separation of variables, solve for the integration constant from data, and be able to reason about when the solution is defined on all of $\mathbb{R}$ (preventing denominator zeros via an inequality condition).

\item {\bf 4.A1:}
The problem deals with multivariable calculus in polar coordinates: sketching a region given by radial and angular bounds and evaluating a double integral of $e^{x^2+y^2}$ over that region. A challenge is to visualize two symmetric angular sectors and to set up the change of variables with the Jacobian $r$, noticing that $e^{x^2+y^2}=e^{r^2}$ simplifies the computation. The student needs to master polar sketching, apply the polar substitution with correct limits and Jacobian, and be able to carry out the radial integral exactly.

\item {\bf 5.A1:}
The problem part deals with planar flux and the divergence (Green's) theorem for two rectangular regions, one depending on a positive parameter $a$. A challenge is to convert a difference of boundary fluxes into an area integral of the constant divergence, keep track of orientation and outward normals, and solve for $a$ from the resulting linear relation. The student needs to recognize when the divergence theorem applies, compute areas and signs correctly, and be able to isolate the parameter from the flux constraint.

\item {\bf 5.A2:}
The problem part deals with a scalar line integral along a parametrized curve $\gamma(t)$ over $[0,1]$. A challenge is to express the integrand along the path, compute arc length via $\|\gamma'(t)\|$, and evaluate the resulting one-variable integral by an effective substitution. The student needs to master parameterization and arc length, set up $\int_\gamma g\,ds$ correctly, and be able to perform the substitution to obtain a closed-form value.

\end{itemize}

\begin{figure}[!t]
\centering
\includegraphics[width=\columnwidth]{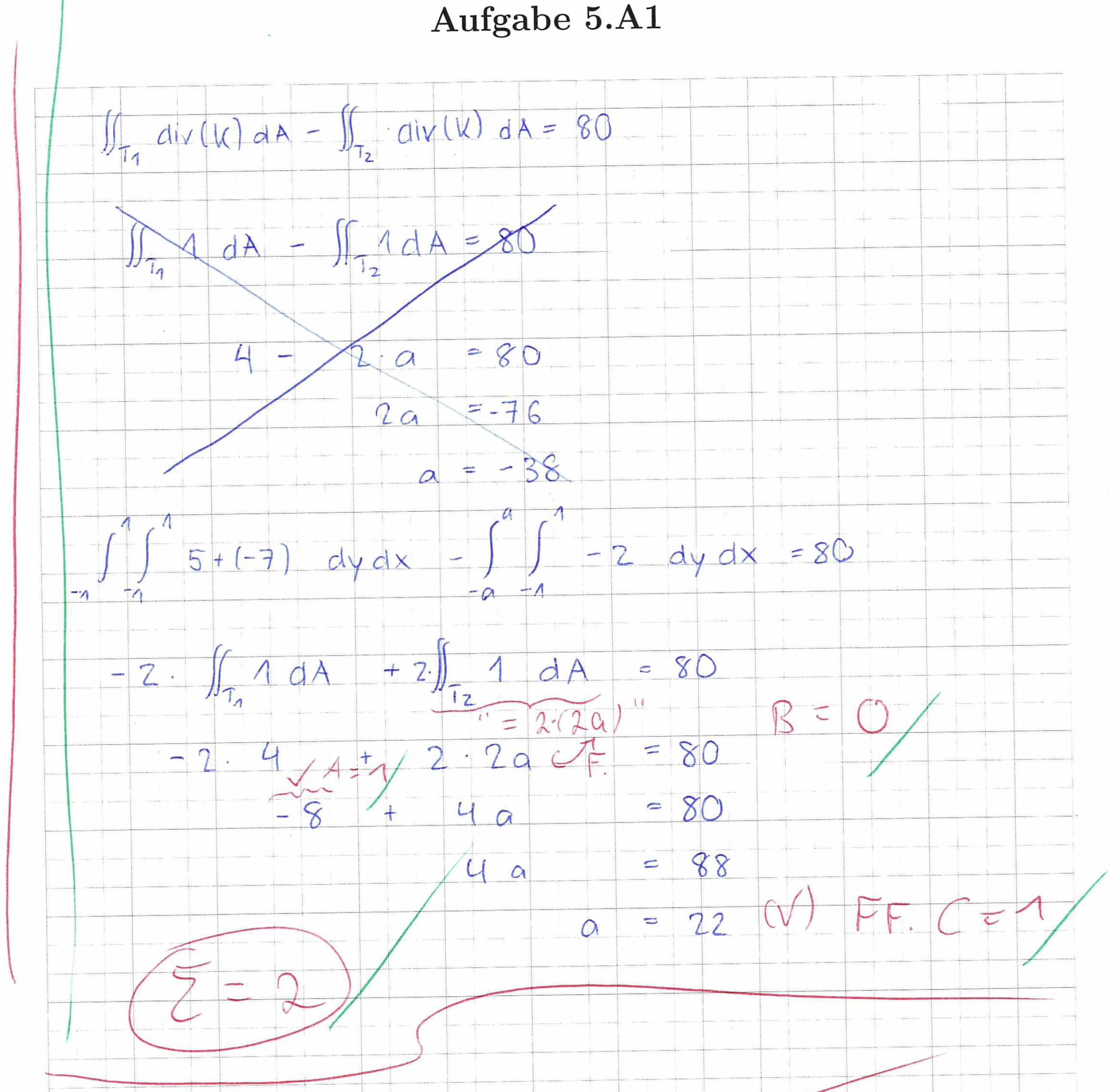}
\caption{An example of graded student work in the answer booklet.}\label{fig:example}
\end{figure}

Students were instructed to write their solutions on 10~blank, labelled pages in a separate answer booklet, where two pages were dedicated each to problems~2 through~4, and two pages each for the two parts of problem~5. These were pre-printed and personalized, so they already included student-identifying information.

\subsection{Sample}
\begin{figure}[!t]
\centering
\includegraphics[width=\columnwidth]{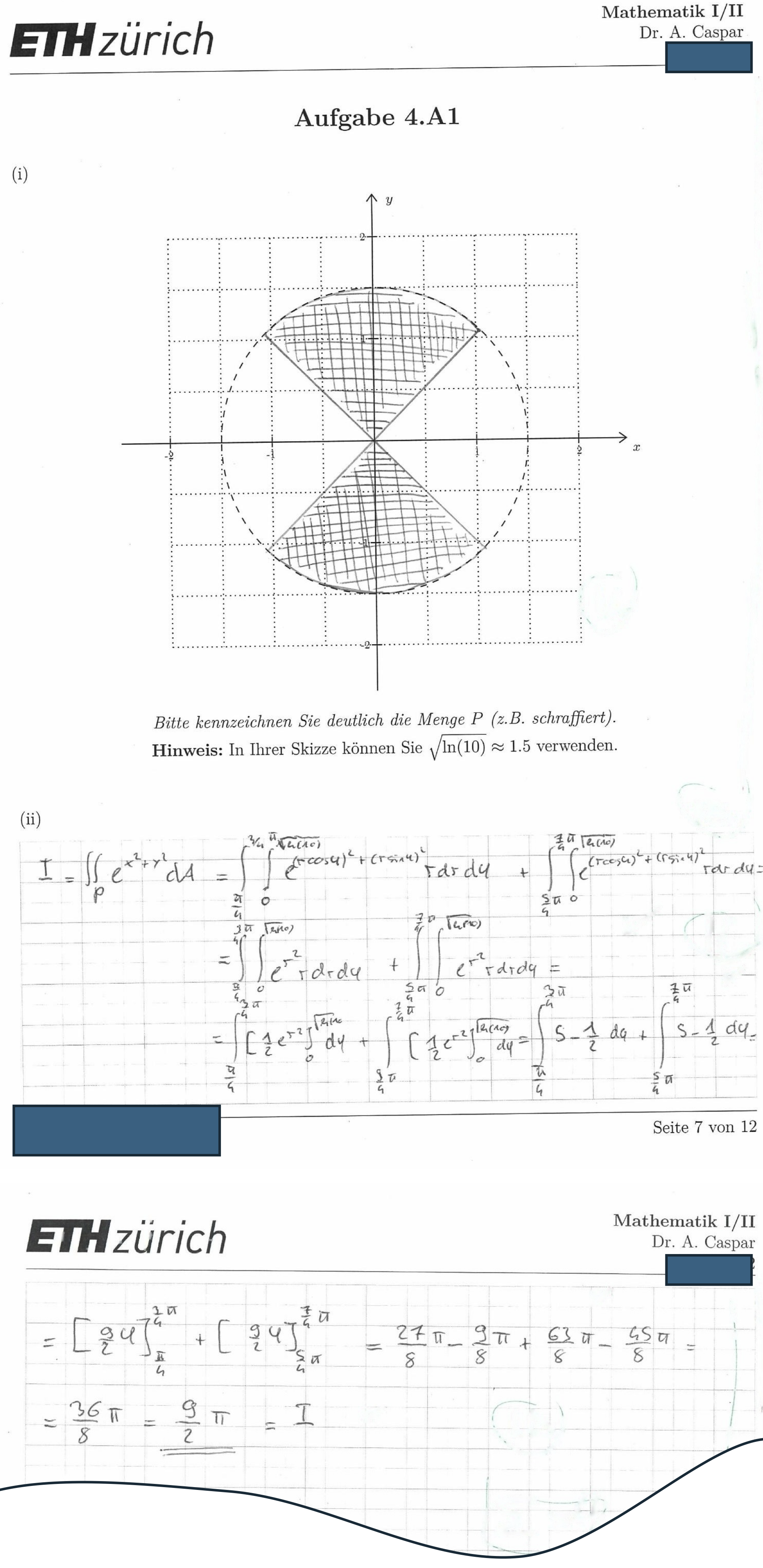}
\caption{An example of the input for the AI-system; grading marks were removed and two pages combined into one image. Potentially identifying information was redacted here for publication purposes (dark blue boxes).}\label{fig:input}
\end{figure}

The student work was graded by teaching assistants (TAs) based on four to six rubric items per problem. For our study, we scanned 349~exams; Figure~\ref{fig:example} shows a short excerpt as an example of graded student work. We were able to remove the red, pink, and green grading marks using the image-editing package OpenCV~\cite{opencv}, as  students were not allowed to use these colors, and we combined the pages for each problem, or, in case of problem~5, problem part; Figure~\ref{fig:input} shows an example.

In parallel, we reentered the $349\mbox{ students}\times19\mbox{ rubric items}=6631\mbox{ TA-grading decisions}$ manually based on the scans, as the original exam spreadsheets only listed their per-problem sums. This established the ground truth for our study, assuming that all TA decisions were correct.

While entering the TA data, we noticed that students did not always follow directions: some students did not write their solutions on the designated pages, others attached extra sheets. To authentically model a grading workflow, we did not clean up these situations, except for five exam booklets where the students did not even attempt to follow any of the guidelines. We also found that due to clerical error, when manually entering the exam numbers for AI grading, two records were assigned non-existing numbers; instead of heuristically fixing this, we discarded those records. This left us with 342~out of the originally 349~exams, which form the matched  dataset for our study.

Another behavior we observed was that a large number of students partially or completely avoided the open-ended questions altogether. This is understandable, given that the majority of exam points could be gained by correctly answering the closed-answer questions; as typical for exams in the German-speaking university tradition, full points were not expected to receive a good grade.

\subsection{AI Grading}

\begin{figure}[!t]
\centering
\includegraphics[width=\columnwidth]{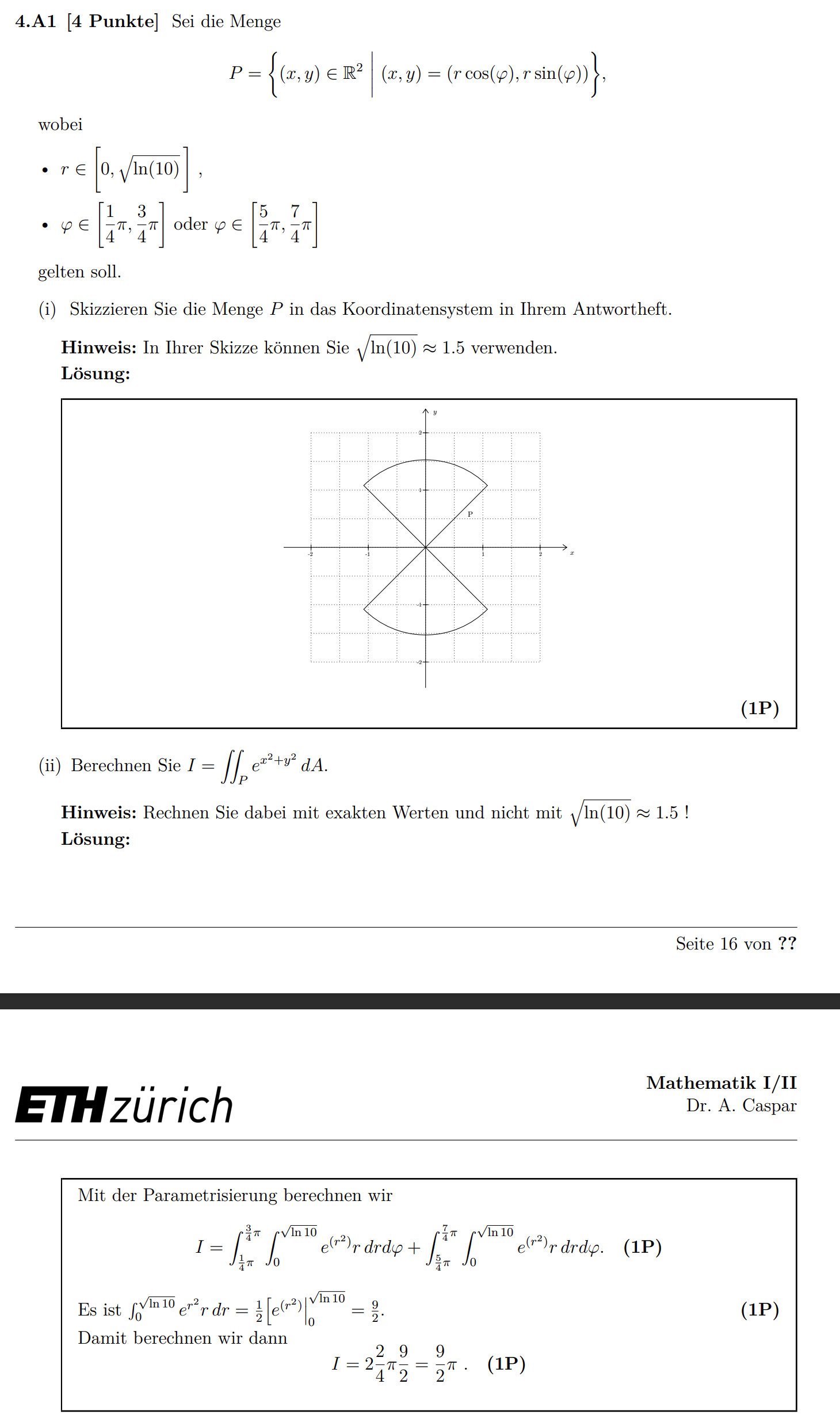}
\caption{An example of the grading rubric, provided in the same format to the TAs and the AI.}\label{fig:rubric}
\end{figure}

The rubric for AI grading was provided in the same form as for the TAs, Fig.~\ref{fig:rubric} shows an example. Rubric items are denoted by the point-value labels (e.g.,~``(1P)'') --- all except one rubric item were worth one point, that item in the last part of problem~5 being worth 2~points.

We accessed OpenAI models via Azure AI Services~\cite{azure} under the university's privacy-preserving contractual framework. Specifically, we used the multimodal reasoning model GPT-5~\cite{gpt5} (model version 2025-08-07) hosted in a Swedish data center. The code, including the prompt, is available at \url{https://gitlab.ethz.ch/ethel/mathexam/-/blob/main/grade.py}.

Each submission consisted of two images: (i) a page of student work (see Fig.~\ref{fig:input}) and (ii) the corresponding rubric page (see Fig.~\ref{fig:rubric}). Grading proceeded strictly page-by-page, one page at a time; loose extra sheets that some students submitted could not be accommodated by this mechanism. To deliver the images, our server issued short-lived, randomized (``ephemeral'') URLs that were valid only for the duration of a single request (two at a time). We avoided embedding the images directly to prevent exceeding the model's context window. In contrast to the TAs, which only gave whole (integer) points, the AI was prompted to also assign partial (fractional) credit. In total, we had $344\times5=1720$ grading cycles, which we were able to evaluate with several agents working in parallel within a little more than seven hours. 

A design goal was to mirror a plausible production workflow. Grading assistance for exams must not require cumbersome data preparation or prohibitive manual effort. For broad, scalable use, the process has to yield a clear net reduction in workload, since otherwise it makes little sense to deploy it beyond possible gains in fairness and objectivity, and it has to be robust and intuitive enough to operate with minimal technical support.

\subsection{Confidence Filters}
Generative AI will always produce an answer, regardless of whether that answer is reliable or not.. Thus, beyond computing AI-based scores, an equally important task is quantifying how much we should trust each score~\cite{kortemeyer2025assessing}. ``Confidence'' here differs from conventional quality metrics: in production use the ground truth is unknown, so confidence must be inferred from information available at decision time (the AI's own outputs and model-based expectations). To do so, we draw on classical and Bayesian statistics.

In production, there is no ground truth available up front: initially, the AI assigns a provisional score to every student-item. A confidence filter then accepts or rejects each AI judgment. The filter's parameters reflect risk tolerance: stricter settings reduce auto-acceptance and increase human workload; looser settings do the opposite.

\subsubsection{Partial-credit Threshold}
Evidence from earlier studies indicates that AI graders tend to be conservative --- more prone to mark correct work as incorrect (false negatives) than the reverse (false positives)~\cite{kortemeyer2025assessing}. A simple, conservative safeguard is therefore to binarize partial credit using a threshold $t$ for ``correctness.'' For filtering purposes, any student-item with AI score below $t$ (i.e., incorrect or ``insufficiently correct'') is flagged for human review, while only items at or above $t$ are considered for auto-acceptance (subject to the risk screen below).

\subsubsection{Risk Threshold}
Modern psychometrics, particularly Item Response Theory (IRT), models student ability and item difficulty as latent variables inferred from observed responses. A fitted IRT model can be used predictively to set expectations for each student-item pair. We employ a two-parameter logistic model to estimate the expected probability of success (or, under partial credit, a normalized expected value) for student~$i$ on item~$j$:
\begin{equation}\label{eq:irt}
p_{ij}=\frac{1}{1+\exp\!\bigl(-a_j(\theta_i-b_j)\bigr)}\,,
\end{equation}
where $\theta_i$ denotes the latent ability of student $i$, and $a_j$ and $b_j$ are the discrimination and difficulty of part $j$, respectively~\cite{rasch,kortemeyer2019quick}.

Let $s_{ij}\in[0,1]$ be the AI's normalized score for the same student-item. We define the \emph{risk} of accepting that AI judgment as the absolute deviation between observed and expected~\cite{sinharay2006posterior}:
\begin{equation}\label{eq:risk}
\text{Risk}_{ij}=\bigl|\,s_{ij}-p_{ij}\bigr|\,.
\end{equation}
Given a tolerance $r\in[0,1]$, we accept the AI decision if $\text{Risk}_{ij}\le r$ and route it to a human grader otherwise. Intuitively, when the AI's score aligns with what the IRT model predicts for that student on that rubric time, the decision is unsurprising and low-risk; when it diverges, we seek human judgment.

In practice, the partial-credit threshold $t$ and risk tolerance $r$ are tuned jointly to meet explicit operating targets (e.g., precision on auto-accepted full-credit decisions, upper bounds on false positives) while managing human workload.

%
%

\section{Results}\label{sec:results}
\subsection{Unfiltered Outcome}

\begin{figure}[!t]
\centering
\includegraphics[width=\columnwidth]{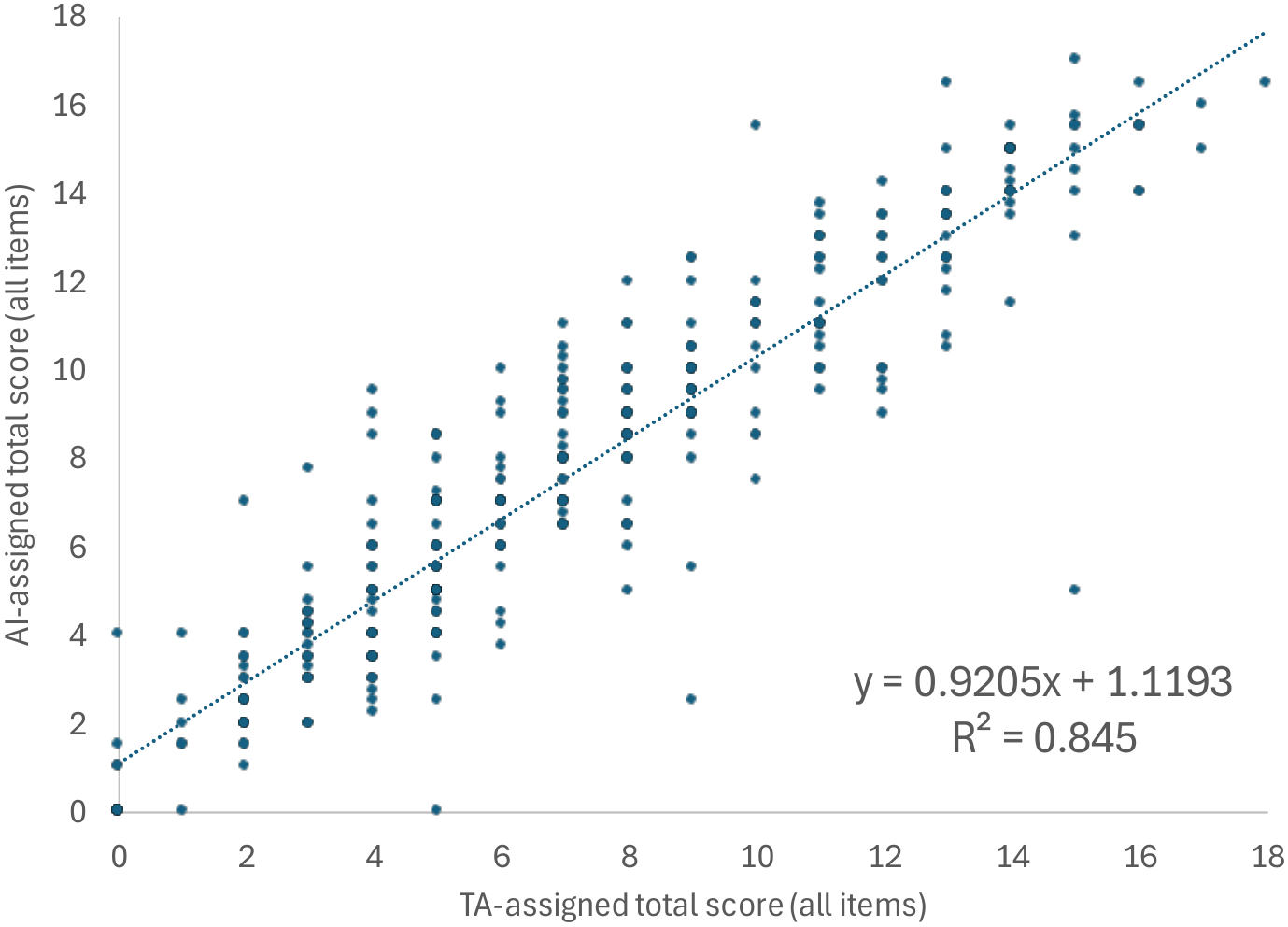}
\caption{Total AI-assigned versus total TA-assigned score. Each data points represents one exam.}\label{fig:merged}
\end{figure}

Figure~\ref{fig:merged} shows the raw grading result: total AI-assigned versus total TA-assigned score without applying confidence filters. While the TAs assigned only integer-point values, the AI was prompted to provide partial credit, resulting in the discontinuous vertical alignment of the data points; overall, though, the AI in all but 8\% of the cases assigned integer points.

The TAs assigned an average total score of $7.35$, compared to $7.89$ for the AI. This appears to indicate that the AI more freely gives away points than the TAs, however, the linear regression between the scores shows that overall, the AI is more conservative (slope $\approx0.92<1.0$), but across the board gives one more point (offset $\approx1.12>0.0$). The coefficient of determination $R^2\approx0.85$ may be sufficient to give feedback on low-stakes formative assessments, but not on high-stakes exams. Also, some extreme outliers are noticeable, underlining the necessity to not blindly trust the AI-results.

\subsection{Filtered Outcomes}
\subsubsection{IRT Analysis}

\begin{figure*}[!t]
\centering
\includegraphics[width=\columnwidth]{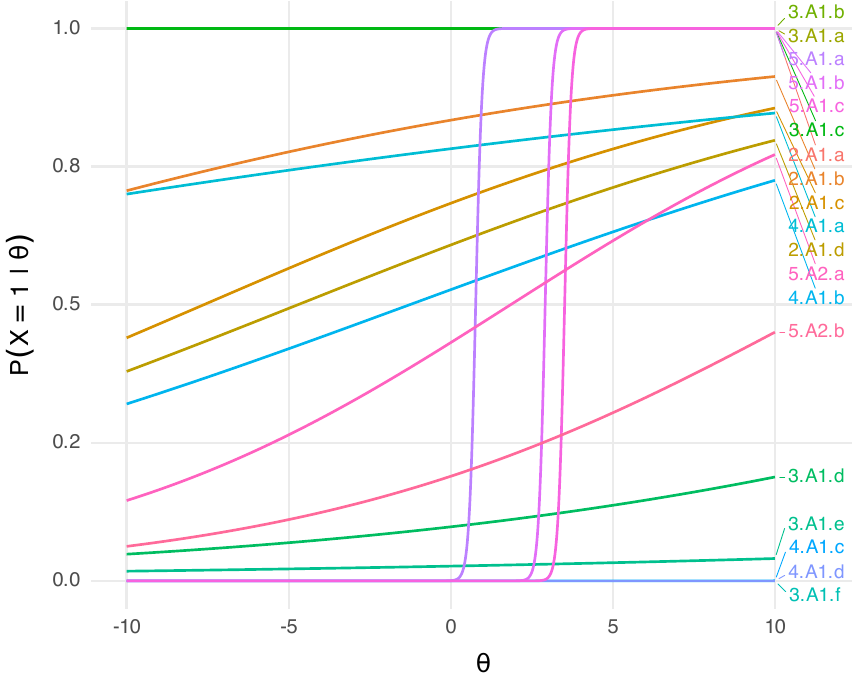}
\includegraphics[width=\columnwidth]{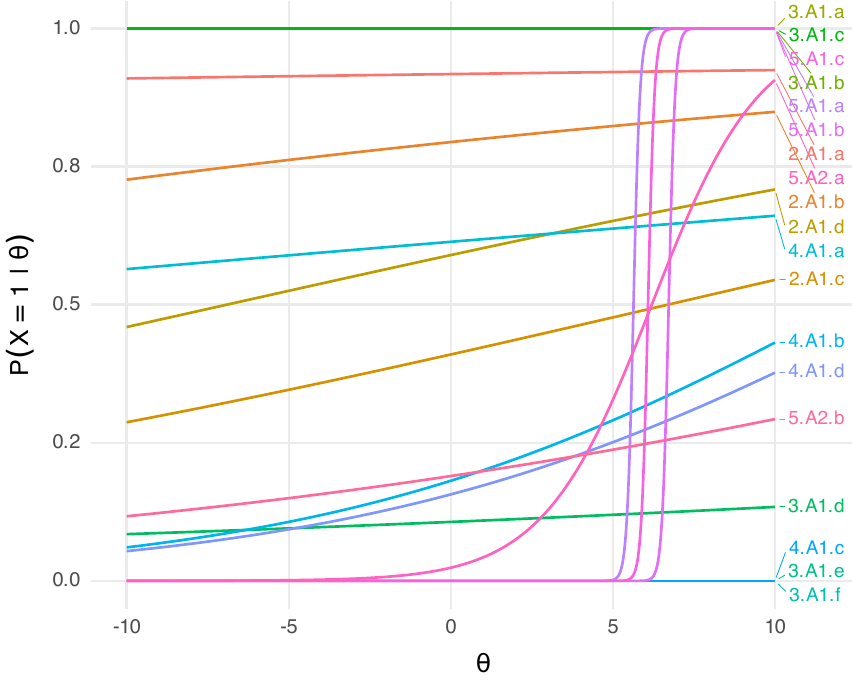}
\caption{Graphs of the function Eq.~\ref{eq:irt}  (item characteristic curves) based on the AI grading (left panel), which will be used for the confidence filtering, and on the TA grading (right panel), given for comparison.}\label{fig:irt}
\end{figure*}

Figure~\ref{fig:irt} shows the probability functions $p_{ij}(\theta_i)$ (Eq.~\ref{eq:irt}) of the rubric items, also known as item characteristic curves, resulting from the IRT estimates. For each rubric item, the left panel shows the likelihood of correctly solving it as a function of the latent ability trait of the students (for example, based on the outcome of the AI grading, a student with estimated ability~$5$ would have a likelihood about $0.3$ to correctly solve item 5.A2.b (bright red curve), the second rubric item of the second part of problem~5. Higher item difficulty shifts these curves to the right toward higher ability, while higher discrimination leads to a steeper transition from low to high success probability as student ability increases. 

In a production setting, no ground truth would be available; for reference, the right panel nevertheless shows the corresponding curves computed from TA grades. With the exception of three items, the item-characteristic curves are largely flat rather than the expected S-shaped functions that rise from near-zero probability at low ability to near-certainty at high ability. In both AI and TA grading, virtually all students correctly solve items 3.A1.a--c, whereas virtually no students correctly solve 3.A1.f and 4.A1.d. Items 5.A1.a--c exhibit extreme discrimination and effectively act as gatekeepers.

Several factors could underlie these mostly undesirable psychometric properties. A likely driver is effort allocation: students appear to have devoted less time and care to open-ended problems than to higher-value multiple-choice questions, which would depress discrimination. In addition, some students may have attempted more difficult-looking open-ended items only when confident and with time remaining, yielding quasi-dichotomous behavior on those parts.

\subsubsection{Risk}
\begin{figure}[!t]
\centering
\includegraphics[width=0.92\columnwidth]{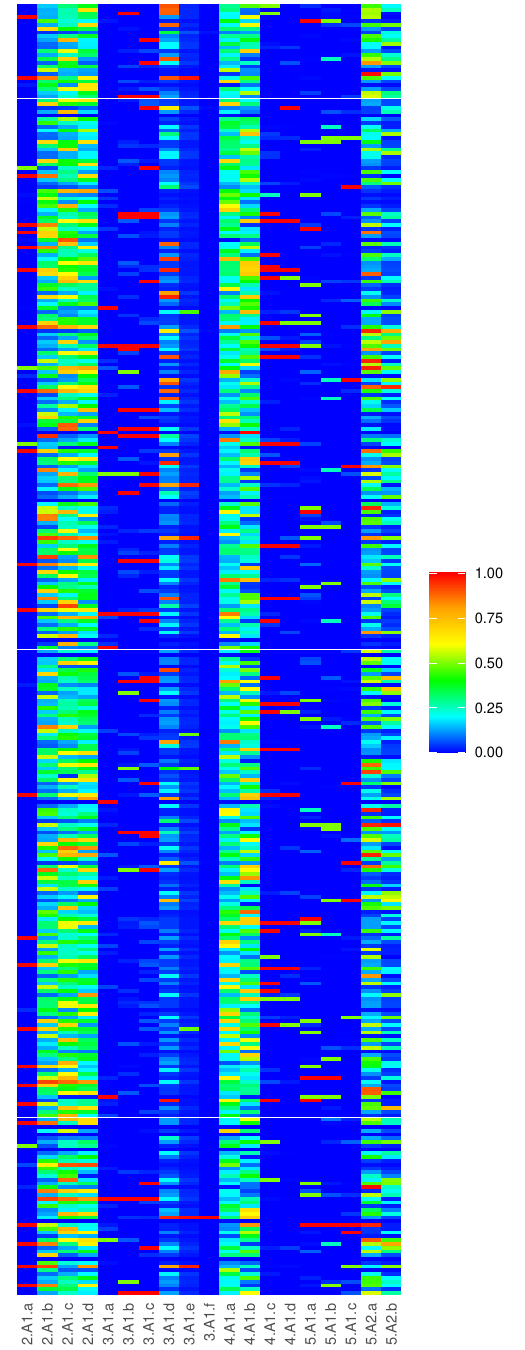}
\caption{Heatmap of Risk (Eq.~\ref{eq:risk}), with the rubric items in its columns and the students in its rows. Blue indicates no risk and red indicates high risk.}\label{fig:heatmap}
\end{figure}

\begin{figure*}[!t]
\centering
\includegraphics[width=0.8\textwidth]{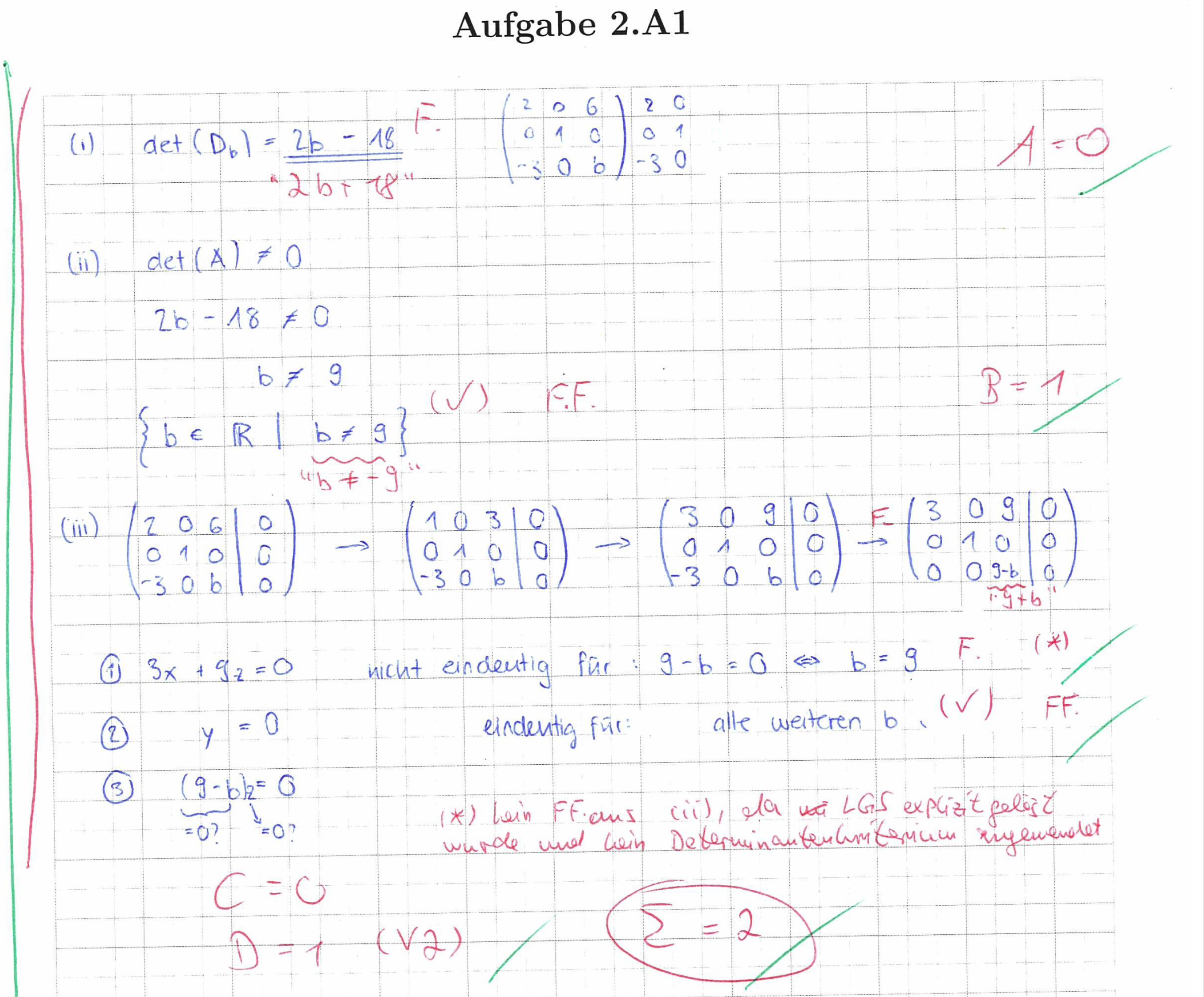}
\caption{Example of student work where four AI-grading decisions in a row were labelled ``high risk''.}\label{fig:followup}
\end{figure*}

Figure~\ref{fig:heatmap} shows a heatmap of $\mbox{Risk}_{ij}$ as defined in Eq.~\ref{eq:risk}; blue indicates perfect agreement between the AI decision and the expected value from IRT, while red indicates that the AI decided to the opposite of the expected value. The columns correspond to the rubric items, and the emerging vertical stripes to items that were graded as expected for nearly all students: the calculation of the determinant, the problem on differential equations, the last two items of the problem on polar coordinates (see Fig.~\ref{fig:rubric}), and the problem part on Green's theorem. The system made more unexpected decisions for the items where green vertical stripes emerge, which includes the graphical task in the problem on polar coordinates. The red lines which emerge for some students are due to various reasons: in some cases it is simply illegible handwriting, but there are also cases where an otherwise high-ability student makes an unexpected error due to an oversight. Figure~\ref{fig:followup} shows an excerpt of an exam where four AI-grading decisions were discarded in a row.

The otherwise high-ability student makes a sign error in the first part of the problem, which was unexpected based on his or her overall performance. The TA nevertheless awarded follow-up points for subsequent rubric items, since they were correctly calculated based on the wrong initial result. The AI graded both the first and the second rubric item as incorrect, being generally unable to award follow-up points, and in total awarded~1 instead of~2 points for the problem. In a production scenario, this problem would be reviewed by a TA based on the risk assessment.

\subsubsection{Balance of Thresholds}
\begin{figure*}[!t]
\centering
\includegraphics[width=\columnwidth]{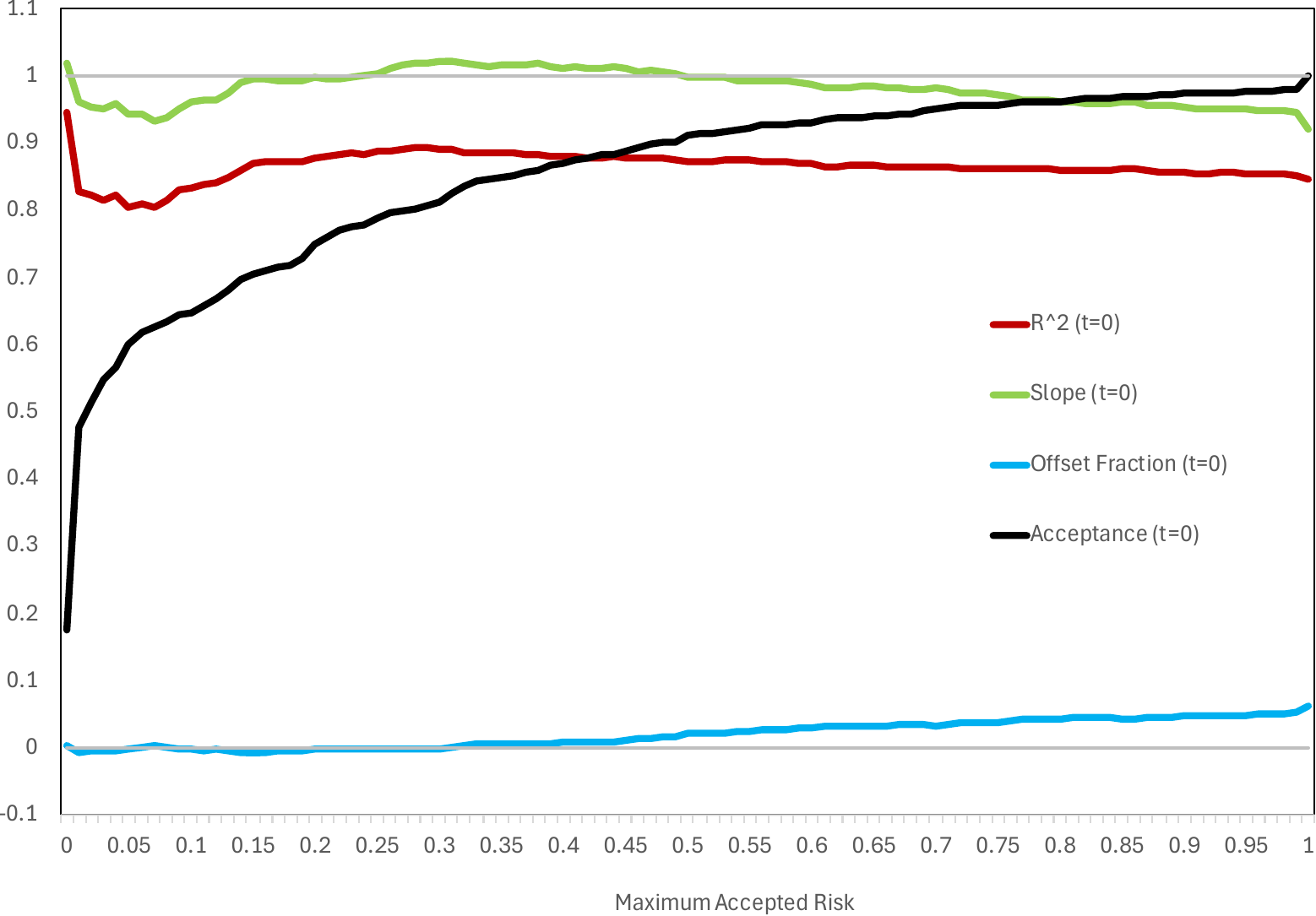}
\includegraphics[width=\columnwidth]{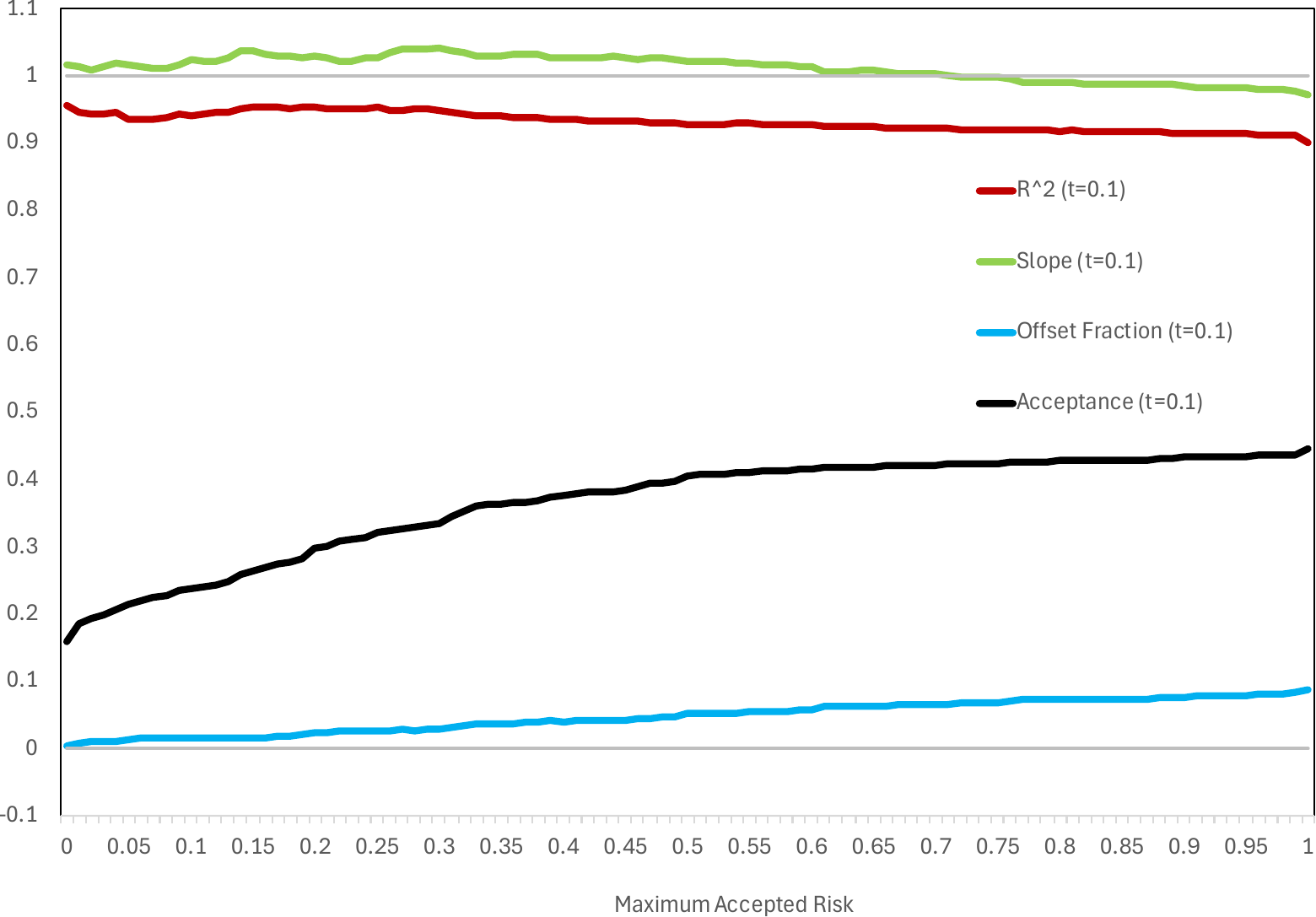}

\includegraphics[width=\columnwidth]{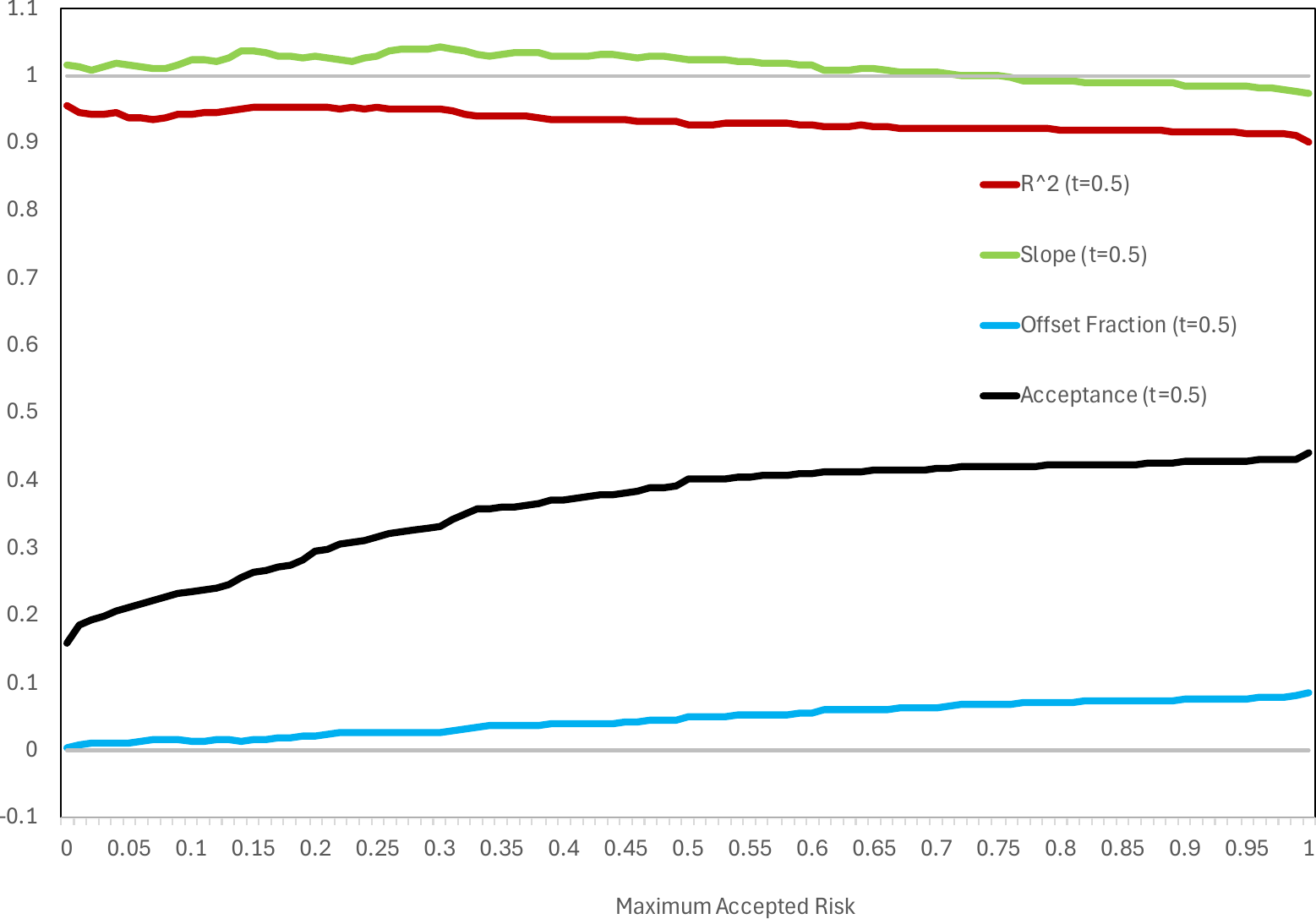}
\includegraphics[width=\columnwidth]{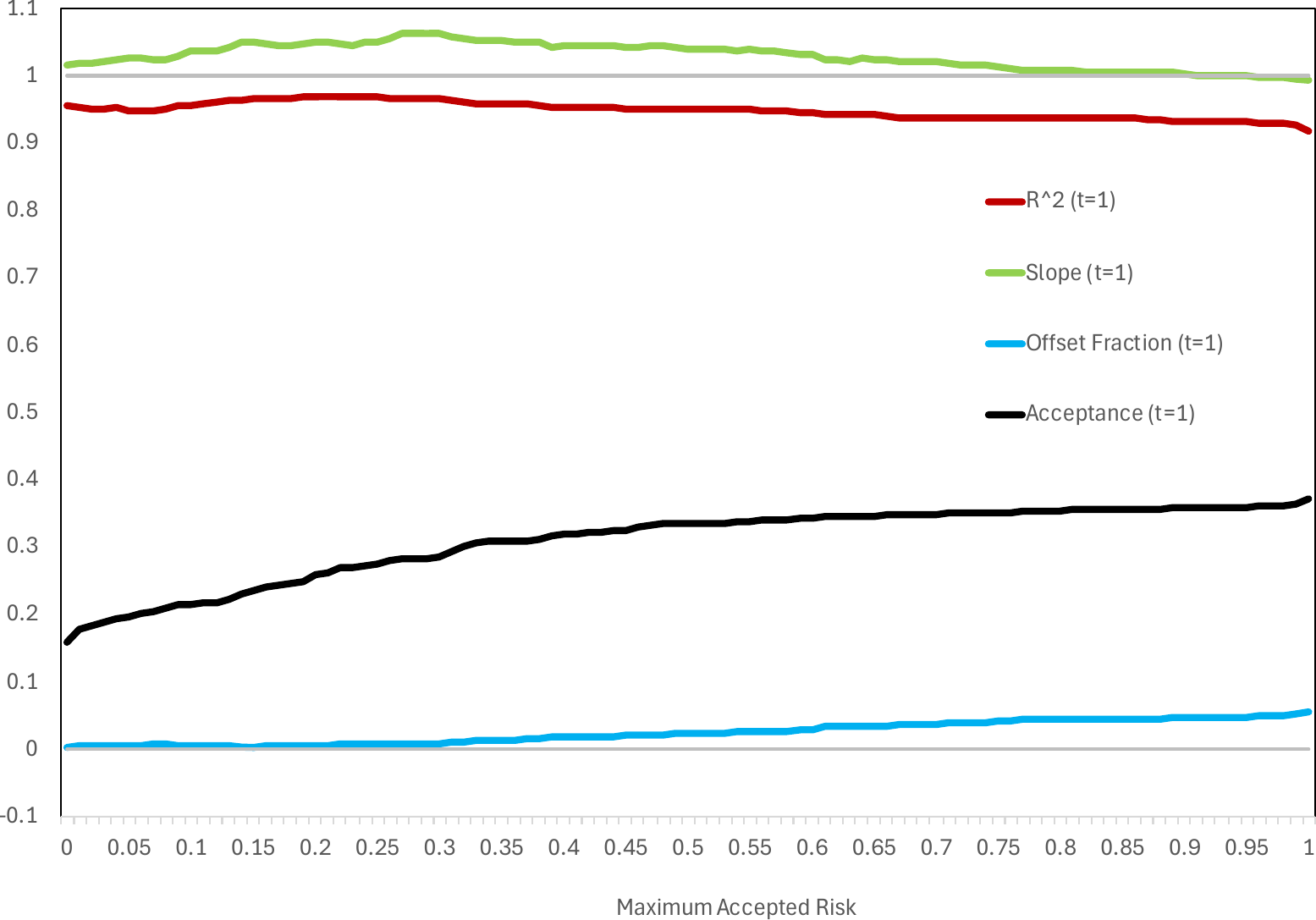}
\caption{Coefficient of determination $R^2$, slope, offset fraction, and acceptance rate as a function of maximum risk threshold $r$  for different values of the minimum partial credit threshold $t$.}\label{fig:riskpartial}
\end{figure*}

Based on the  risks $\mathrm{Risk}_{ij}$, we examine how varying the maximum risk threshold $r$ interacts with different minimum AI partial-credit thresholds $t$. Figure~\ref{fig:riskpartial} reports, for varying $(r,t)$, the coefficient of determination ($R^2$), the slope, the normalized intercept ($\mbox{offset fraction}=\mbox{offset}/18$), and the acceptance rate, that is, the proportion of student-items automatically graded (i.e., passing both thresholds).

Choosing $(r,t)$ is therefore a balancing act between grading accuracy and the manual workload created by rejected AI decisions. For example:
\begin{itemize}
  \item With no minimum partial-credit threshold ($t=0$), a risk cap of $r=0.3$ yields $R^2\approx 0.89$, $\mathrm{slope}\approx 1.02$, $\mathrm{offset\ fraction}\approx 0.00$, and an acceptance rate of $81\%$. In other words, roughly one fifth of student-items would require manual grading.
  \item With a mild partial-credit threshold ($t=0.1$; discarding only items that received essentially no AI credit), at $r=0.2$ we obtain $R^2\approx 0.95$, $\mathrm{slope}\approx 1.03$, and $\mathrm{offset\ fraction}\approx 0.02$ --- an almost perfect fit, but at the cost of manually grading about $70\%$ of the student-items. The exact value of the partial-credit threshold does not have much influence on the outcome beyond choosing $t=0$ and $t>0$, as only 8\% of the AI-grading decisions resulted in partial credit --- essentially, $t>0$ just filters out all student-items that the AI graded as ``wrong,'' with not much distinction between ``degrees of wrongness.''
\end{itemize}

Filtering the AI-grading decisions yields better fits than the unfiltered results (Fig.~\ref{fig:merged}; $R^2\approx0.85$,  slope$\approx0.92$, $\mathrm{offset\ fraction}\approx0.06$), but far less than a 100\% acceptance rate.

\subsection{Observations}\label{sec:observations}
Three contextual factors of this deployment plausibly shaped the observed psychometrics and the behavior of the confidence filter, independent of the mathematical quality of the exam itself.

First, the open-ended portion constituted a small share of the total points and, anecdotally from the scans, attracted uneven student effort relative to the high-value multiple-choice section. This is consistent with the largely flat item-characteristic curves in Fig.~\ref{fig:irt} and the near-ceiling/near-floor behavior on several rubric items. Such patterns limit discrimination not because the items are poorly written, but because many students either dispatched the easiest open-ended steps quickly or --- under time pressure --- attempted only selected harder parts. In this regime, even a well-calibrated filter cannot recover strong ability gradients.

Second, the analysis operated on a small set of indicators (18~rubric items across five problem prompts). With so few observable ``slots,'' IRT parameter estimates and downstream risk (Eq.~\ref{eq:risk}) become sensitive to idiosyncrasies in response patterns. This helps explain the pronounced vertical striping in Fig.~\ref{fig:heatmap}: information concentrates in a handful of rubric elements, while others contribute little variance. More items, or more granular rubric checkpoints, generally stabilize discrimination estimates and yield a smoother trade-off curve in Fig.~\ref{fig:riskpartial}.

Third, several layout and workflow realities reduced effective observability. A non-trivial number of students wrote outside the designated regions or appended loose sheets. Because grading proceeded page-by-page with one combined image per prompt, work placed on the ``wrong'' page could be missed, and the model's judgments would then reflect absence of evidence rather than evidence of absence. This is a recognizable failure mode in mixed-media grading and likely accounts for some of the red bands by student in Fig.~\ref{fig:heatmap}.

\begin{figure*}[!t]
\centering
\includegraphics[width=\columnwidth]{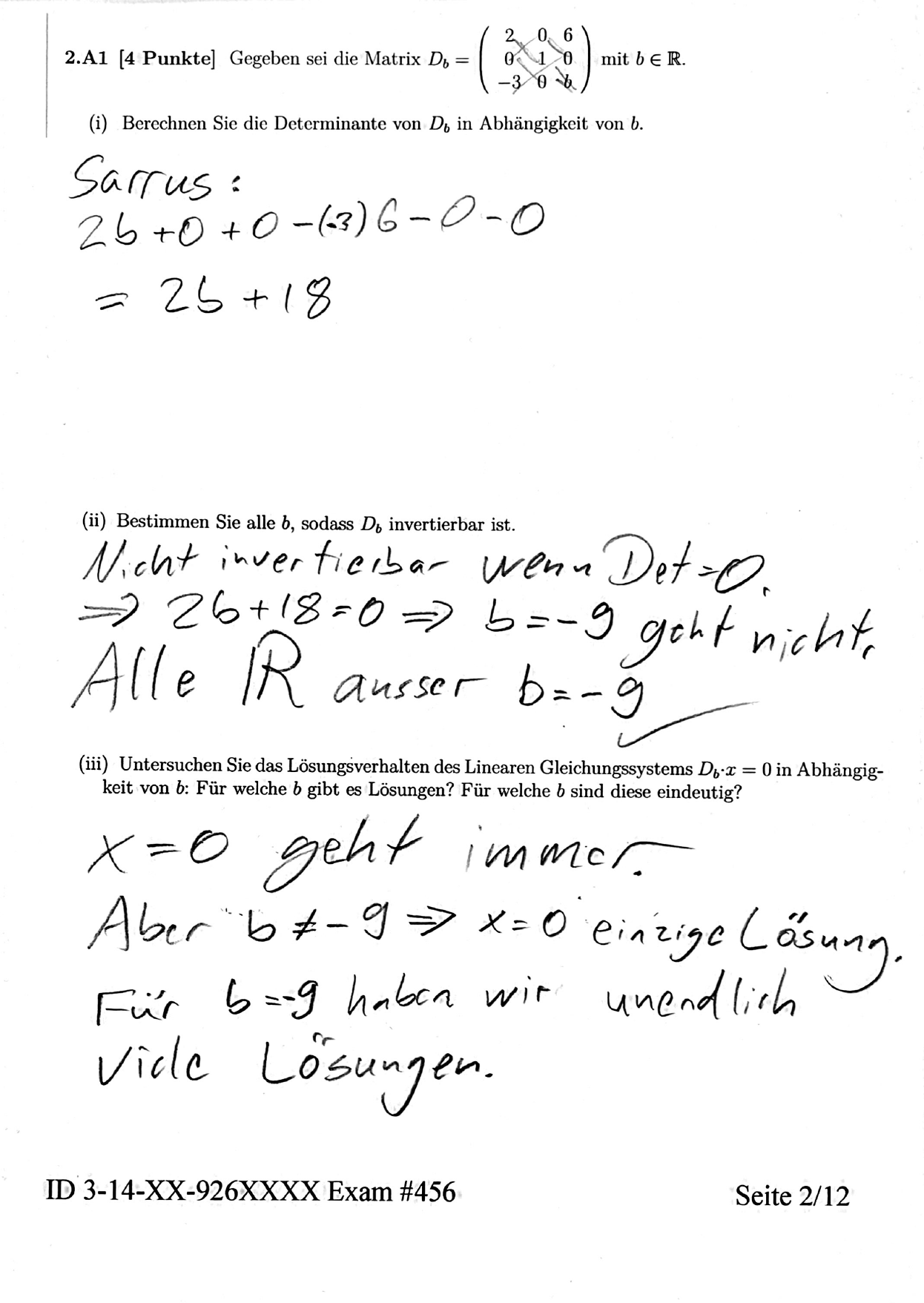}
\includegraphics[width=\columnwidth]{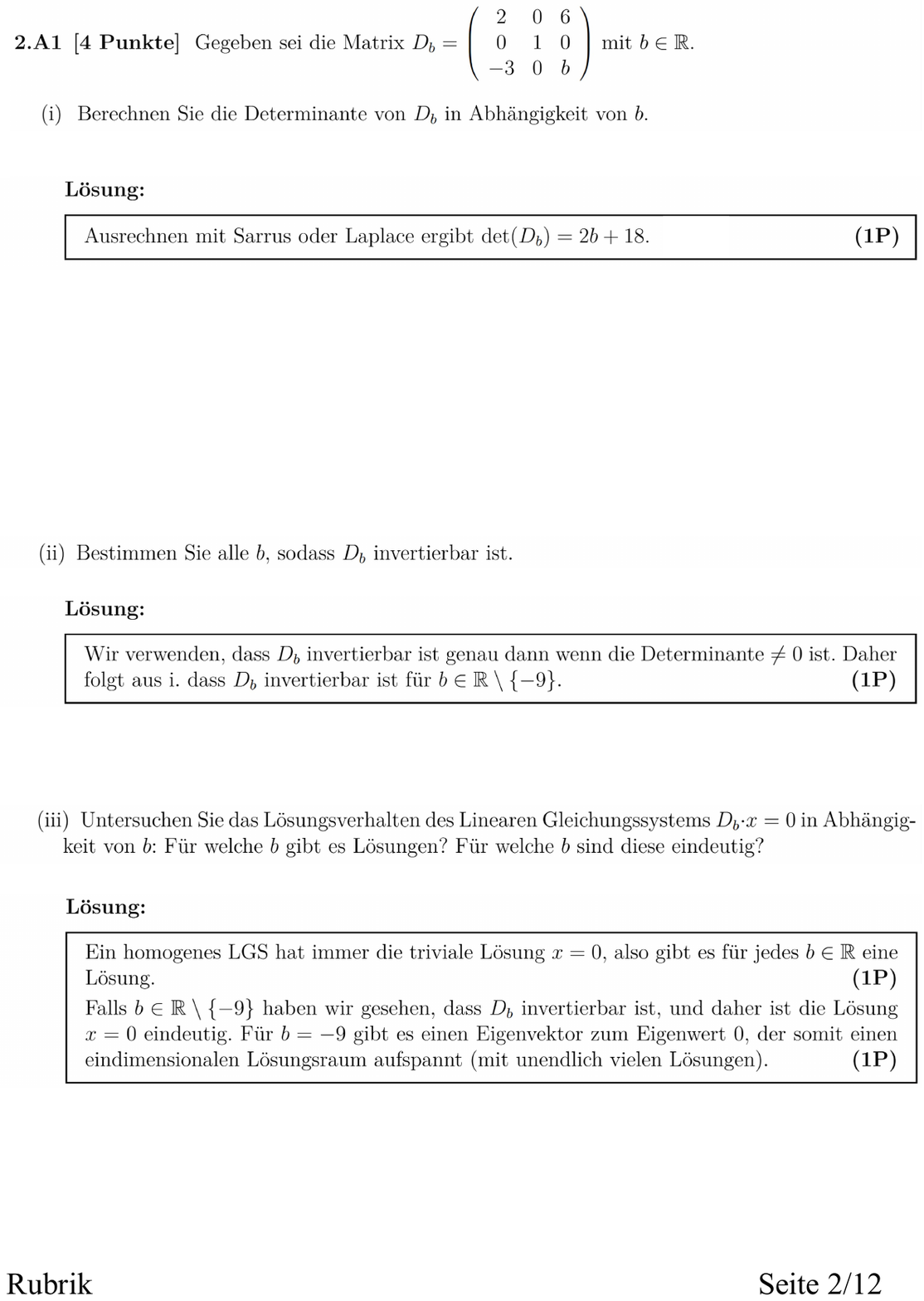}
\caption{Mockup of a reliable layout of question sheet (left panel) and rubric (right panel).}\label{fig:layout}
\end{figure*}

\section{Discussion}\label{sec:discussion}
The results are moderately promising but not perfect. Production-ready results can only be achieved for about 30\% of the grading load, which is less than in our earlier studies of physics~\cite{kortemeyer2025assessing} and chemistry exams~\cite{cvengros2025assisting}.
This might be improved by some practical, incremental adjustments that do not alter the mathematical substance of the exam:
\begin{itemize}
\item \emph{Assessment weight and time budgeting.} Ensure that open-ended items contribute sufficient points and protected time to elicit consistent effort; even modest re-weighting can restore discrimination without changing content.
\item \emph{Increase observable checkpoints.} Where pedagogically sensible, split multi-step solutions into two or three rubric-visible substeps (clear, independent criteria). This raises the effective item count and improves IRT stability, but of course has to be balanced against the desired open-ended character of the tasks. 
\item \emph{Stronger spatial anchoring.} Use clearly designated answer regions with brief labels that mirror rubric keys (e.g., ``4.A1.a: sketch''), and include page anchors/registration marks. Alternatively, the questions can be directly on the answer sheet, with sufficient space even for meandering answers, so the student work can mirror the rubric. Instruct proctors to remind students to keep work within the designated areas. Figure~\ref{fig:layout} shows a mockup  of how student work and rubric would ideally align.
\item \emph{Cleanliness.} Background grids are generally problematic, as they interfere with the OCR. Also, students should be asked to use pencil and erasers rather than crossing out solution attempts. As the first step in processing the solutions is scanning them, and in future workflows, grading results can likely be viewed online, having non-permanent markers does not invite retroactive manipulation.
\end{itemize}

The exam was mathematically sound and well structured for the course, while several extrinsic factors --- relative stakes, indicator count, and spatial discipline of responses --- constrained what psychometrics and automation could extract. The recommended adjustments target those constraints and, if adopted, should improve both calibration and acceptance rates at fixed quality targets without diluting the assessment of authentic mathematical reasoning.

\section{Conclusion}\label{sec:conclusions}
Our study demonstrates that calibrated, human-in-the-loop use of contemporary multimodal LLMs can shoulder a meaningful share of grading for open-ended calculus work without eroding the evidentiary value of students' reasoning. Unfiltered, AI-TA agreement was moderate ($R^2\approx0.85$, slope $\approx0.92$ with a positive offset), which is adequate for low-stakes feedback but not for high-stakes decisions. Confidence filtering that combines a partial-credit screen with an IRT-based risk test improved agreement substantially while making the workload-quality trade-off explicit: with no partial-credit floor and a risk cap of $r!=!0.3$, the system auto-accepted about $81\%$ of student-items at $R^2\approx0.89$; under stricter settings ($t!=!0.1$, $r!=!0.2$) agreement rose to $R^2\approx0.95$ at the cost of manual review for roughly $70\%$ of items. In short, the pipeline can deliver production-ready decisions for a sizable subset of routine cases, provided that ambiguous or low-signal cases are routed to experts via conservative operating points.

The deployment also clarifies where incremental design choices will raise ceiling performance. Three factors constrained psychometric leverage here
limited weight on open-ended tasks, a small set of rubric checkpoints, and occasional misalignment between designated answer regions and where work actually appeared. None of these implicate the mathematical quality of the exam, and all admit pragmatic remedies: modestly increasing the contribution and protected time for open-ended items, adding a few rubric-visible substeps where pedagogically natural, strengthening spatial anchoring and multi-page capture, and pooling anchor items across terms to stabilize calibration.  Taken together, the results support a modest but optimistic conclusion: with calibrated confidence and simple layout affordances, AI can make grading of authentic calculus work more scalable while reserving human judgment for the cases where it matters most.

\backmatter

\bmhead{Acknowledgements}

We would like to thank Maike Tauschhuber for her assistance in data collection. We would also like to thank Anna Kortemeyer for proofreading the manuscript.

\section*{Declarations}

\subsection*{Funding}
Not applicable.

\subsection*{Conflict of interest/Competing interests}
Not applicable.

\section*{Ethical Considerations}
This study was approved by the ETH Zurich Ethics Commission (IRB) as protocol EK 2023-N-366.

\section*{Consent to Participate}
Consent to participate was not obtained for historical exam data, which was obtained within the framework of regular instruction (no experimental procedure). Per approved IRB protocol,  data was depersonalized using a double-blind mechanism between original course instructors and staff carrying out the analysis.

\section*{Consent to Publish}
Not applicable.

\subsection*{Data Availability}
In adherence to the IRB protocol, data is not available.

\subsection*{Materials Availability}
Not applicable.

\subsection*{Author Contribution}
Gerd Kortemeyer had the overall project lead and drafted the first version of the manuscript. Alexander Caspar as course instructor accompanied all aspects of the study. Daria Horica was involved in data collection, data cleaning, and data preparation. All authors collaborated on the submitted version of the manuscript.



\bibliography{references}

\end{document}